\documentclass[12pt,a4paper]{article}
\usepackage{amsmath}
\usepackage{amssymb}
\usepackage{bm}
\usepackage{color}

\usepackage{hyperref}
\usepackage{graphicx}
\catcode `@=11 \@addtoreset{equation}{section} \catcode `@=12
\def\Tr{{\rm Tr}}

\begin{document}
\setlength{\oddsidemargin}{0cm}
\setlength{\baselineskip}{7mm}

\begin{titlepage}

	\begin{center}
		{\LARGE
		Bootstrap Method in Harmonic Oscillator
		}
	\end{center}
	\vspace{0.2cm}
	\baselineskip 18pt 
	\renewcommand{\thefootnote}{\fnsymbol{footnote}}

	\begin{center}

		Yu {\sc Aikawa}$^{a}$\footnote{%
		E-mail address: aikawa.yu.17(at)shizuoka.ac.jp}, 
		Takeshi {\sc Morita}$^{a,b}$\footnote{%
			E-mail address: morita.takeshi(at)shizuoka.ac.jp
		} and
		Kota {\sc Yoshimura}$^{c}$\footnote{%
		E-mail address: kyoshimu(at)nd.edu
	}

		\renewcommand{\thefootnote}{\arabic{footnote}}
		\setcounter{footnote}{0}
		
		\vspace{0.4cm}
		
		{\it
			a. Department of Physics,
			Shizuoka University \\
			836 Ohya, Suruga-ku, Shizuoka 422-8529, Japan 
			\vspace{0.2cm}
			\\
			b. Graduate School of Science and Technology, Shizuoka University\\
			836 Ohya, Suruga-ku, Shizuoka 422-8529, Japan
			\vspace{0.2cm}
			\\
			c. Department of Physics, University of Notre Dame \\
			Notre Dame, Indiana, 46556, USA
			}

	\end{center}
	
	%\newpage
	
	\vspace{1.5cm}
	
	\begin{abstract}

	\end{abstract}
	%\\

Recently, an application of the numerical bootstrap method to quantum mechanics was proposed, and it successfully reproduces the eigenstates of various systems.
However, it is unclear why this method works.
In order to understand this question, we study the bootstrap method in harmonic oscillators. 
We find that the problem reduces to the Dirac's ladder operator problem and is
exactly solvable analytically.
Our result suggests that the bootstrap method may be regarded as a numerical version of the Dirac's approach and it may explain why it works in various systems.
	
\end{titlepage}

\section{Introduction}
Recently, the bootstrap analysis in zero \cite{Anderson:2016rcw, Lin:2020mme} and one-dimensional systems \cite{Han:2020bkb} have been proposed, and
they are actively studied in various models \cite{Kazakov:2021lel, Berenstein:2021dyf, Bhattacharya:2021btd, Aikawa:2021eai, Berenstein:2021loy, Tchoumakov:2021mnh}.
This method works even in matrix models at $N=\infty$, which is not possible in Monte-Carlo computations \cite{Anderson:2016rcw, Lin:2020mme, Han:2020bkb, Kazakov:2021lel}.
Besides, the method may work even if the Euclidean action contains imaginary terms which cause sign problems \cite{Aikawa:2021eai}.
Hence, this method might play a complementary role to the Monte-Carlo computations.

However, one important question has not been answered, {``\it why does this method works?"}
Besides, the previous results indicate that the method in quantum mechanics works better for deriving observables in lower energy eigenstates than higher ones, but the reason is unclear. 

In this paper, in order to understand these questions, we study harmonic oscillators in one-dimensional quantum mechanics by using the bootstrap method \cite{Han:2020bkb,Berenstein:2021dyf}.
Since harmonic oscillators are simple, it might provide us an insight how the bootstrap method works in quantum mechanics. 
Not only that, since harmonic oscillators are essential building blocks of various quantum systems including quantum field theories, studying harmonic oscillators by using the bootstrap method may be valuable.

We will see that the bootstrap problem in harmonic oscillators reduces to the Dirac's ladder operator problem and is exactly solvable analytically.
This result suggests that the bootstrap method might be regarded as a generalization of the ladder operator method, and it might explain why the numerical bootstrap method works in various systems \cite{Han:2020bkb, Berenstein:2021dyf, Bhattacharya:2021btd, Aikawa:2021eai, Berenstein:2021loy, Tchoumakov:2021mnh}.
We also find that the eigenstates are obtained one by one starting from the ground state.
This explains why the method works better for the lower energy eigenstates.

We also investigate an anharmonic oscillator, which has been studied in \cite{Han:2020bkb, Bhattacharya:2021btd}, by using a bootstrap matrix constructed by ladder operators.
We find that it improves the numerical results.
This also implies the connection between the bootstrap method and the Dirac's ladder operator problem.

The organization of this article is as follows.
In section \ref{sec-review}, we review the bootstrap method in quantum mechanics.
Particularly, we focus on an one-dimensional anharmonic oscillator.
In section \ref{sec-HO}, we apply the bootstrap method to a harmonic oscillator.
We see that the bootstrap problem is analytically solvable.
We also propose an improvement of the bootstrap analysis in the anharmonic oscillator by using ladder operators.
Section \ref{sec-discussion} contains conclusions and discussions.

Note that throughout our paper, we take $\hbar=1$.

%Besides, we find that the exact results in harmonic oscillators are derived even through the numerical bootstrap method.
%This is quite contrasting property compared to other numerical methods such as Monte-Carlo computations, which cannot derive exact results even in simple systems.
%Our result implies that the numerical bootstrap method can distinguish solvable systems.
%As an application, we propose a method to detect unknown solvable systems numerically.

\paragraph{Note}
When we are finalizing this work, a related study appeared \cite{Berenstein:2021dyf}.
There, a harmonic oscillator was solved by using the numerical bootstrap method.
Since they chose different operators in the bootstrap matrix, the exact result was not found.
See Sec.~\ref{sec-numerics} for the details.

\section{Review  of  the bootstrap method in quantum mechanics}
\label{sec-review}

We first briefly review the bootstrap method in quantum mechanics proposed by Han et al.\cite{Han:2020bkb}.

Let us consider an one-dimensional quantum mechanics with a Hamiltonian $H=H(X,P)$.
The idea of the bootstrap method is deriving the spectrum of this system from the positivities of some selected observables.

Suppose that we take $K$ operators $\{ O_n \}$, ($n=1,\cdots,K$). For example $O_1=X$, $O_2=P$ and so on.
Then we define the following operator from them,
\begin{align}
	\tilde{O}=  \sum_{n=1}^{K} c_{n} O_n,
	\label{ops-sample}
\end{align}
where $\{ c_{n} \}$ are some constants.
Since $ \langle \alpha | O^\dagger O  | \alpha  \rangle \ge 0$ is satisfied for any state $|\alpha \rangle $ in this system for arbitrary well-defined operators $O$, 
\begin{align}
	\langle \alpha | \tilde{O}^\dagger \tilde{O} | \alpha  \rangle \ge 0
	\label{positive-cond}
\end{align}
is satisfied too for any constants $\{ c_{n} \}$.
Hence, the following $K\times K$ matrix ${\mathcal M}$ has to be positive-semidefinite \cite{Han:2020bkb},
\begin{align}
	{\mathcal M}:=
	\begin{pmatrix}
		\left\langle O_1^\dagger O_1 \right\rangle & \left\langle O_1^\dagger O_2 \right\rangle     & \cdots & \left\langle O_1^\dagger O_K \right\rangle   \\
		\left\langle O_2^\dagger O_1 \right\rangle &  \left\langle O_2^\dagger O_2 \right\rangle &  \cdots & \left\langle O_2^\dagger O_K \right\rangle   \\
			\vdots & \vdots   & \ddots & \vdots  \\
			\left\langle O_K^\dagger O_1 \right\rangle
		& \left\langle O_K^\dagger O_2 \right\rangle
		& \cdots & \left\langle O_K^\dagger O_K \right\rangle
	\end{pmatrix}
	\succeq 0.
	\label{bootstrap}
\end{align}
Here we have omitted $\alpha$.
This strongly constrains possible expectation values of the operators.
We call ${\mathcal M}$ as a bootstrap matrix.
Note that, as $K$ increases, the constraint would become stronger.

From now, we focus on an energy eigenstate with an energy eigenvalue $E$,
and we take $| \alpha \rangle = | E \rangle  $.
Then, the energy eigenstate has to satisfy the following two additional constraints,
\begin{align}
	&\langle E| \left[ H, O \right] | E  \rangle =0 ,
	\label{HO=0} \\
	&\langle E| HO | E \rangle =E \langle E| O | E  \rangle,
	\label{HO=EO}
\end{align}
for any well-defined operator $O$.
We survey the possible values of $E$ that are consistent with these constraints and the condition ${\mathcal M} \succeq 0$.
If the constraints are sufficiently strong, the ranges of $E$ that satisfies the constraints are highly restricted and may become point-like corresponding to the energy eigenvalues.
In this way, we may obtain the energy eigenvalues in the bootstrap problem.

As a demonstration, let us consider an anharmonic oscillator, which has been studied in Han et al. \cite{Han:2020bkb},
\begin{align}
	H=\frac{1}{2}P^2+\frac{1}{2}X^2+\frac{1}{4}X^4.
	\label{H-AHO}
	\end{align}
In order to construct the bootstrap matrix \eqref{bootstrap}, following Han et al. \cite{Han:2020bkb}, we take the operators
$\{ X^n \}$, $(n=0,1,\cdots,K)$ and define
\begin{align}
	\tilde{O}_X :=  \sum_{n=0}^{K} c_{n} X^n.
	\label{operators-X}
\end{align}
Then, the bootstrap matrix \eqref{bootstrap} becomes $({\mathcal M})_{mn}= \langle X^{m+n-2} \rangle$.
Next we consider the constraints \eqref{HO=0} and \eqref{HO=EO}.
As we show in \eqref{recurrence-general} in Appendix \ref{app-recurrence}, these constraints in this model lead to the recurrence relation
\begin{align}
	n(n-1)(n-2)\langle X^{n-3} \rangle+8En \langle X^{n-1} \rangle-4(n+1)\langle X^{n+1} \rangle-2(n+2)\langle X^{n+3} \rangle=0.
	\label{AHO-recurrence}
\end{align}
By using the parity symmetry of the Hamiltonian, we impose $\langle X^n \rangle=0$ when $n$ is odd.
Then this recurrence relation implies that all expectation values $\langle X^n \rangle$ can be described by $\langle X^2 \rangle$ and $E$.
Now the bootstrap matrix  $({\mathcal M})_{mn}= \langle X^{m+n-2} \rangle$ becomes,
\begin{align}
	{\mathcal M}=
	\begin{pmatrix}
		1 & 0  &  \langle X^2 \rangle  & \cdots  \\
		0 &  \langle X^2 \rangle & 0 &  \cdots  \\
		\langle X^2 \rangle  &  0 & \frac{4}{3} E+ \frac{4}{3} \langle X^2 \rangle   & \cdots   \\
		\vdots & \vdots  & \vdots  & \ddots \\
	\end{pmatrix}. 
	\label{bootstrap-AHO}
\end{align}
We can numerically survey the possible values of $\langle X^2 \rangle$ and $E$ that are consistent with the constraint ${\mathcal M} \succeq 0$.
The results up to $K=11$ are summarized in Table \ref{Table-AHO} \footnote{
    The bootstrap matrix \eqref{bootstrap-AHO} linearly depends on $\langle X^2 \rangle$  while non-linearly depends on $E$.
    Thus, when we fix $E$, the problem finding the minimum (or maximum) of $\langle X^2 \rangle$ that satisfies ${\mathcal M} \succeq 0$ reduces to so called ``Semidefinite Programming Problem", and the numerical costs drastically decrease.
	More specifically, we assign $E$ a number, and try to find the maximum and minimum of $\langle X^2 \rangle$. If the maximum and minimum exist, this number for $E$ is possible. By repeating this procedure for various $E$, we obtain the energy bands, in which $E$ is consistent with ${\mathcal M} \succeq 0$.
    We use Mathematica package SemidefiniteOptimization in our numerical bootstrap analysis.
}.

\begin{table}
	\centering
	\begin{tabular}{|c||c|c|c|c|}%%%The number of columns has to be defined here
	\hline
	 & $n= 1$   &  $n=2$  & $n= 3$  & $n= 4$ \\ %%%% Table body
	\hline
	\hline
	$K=6$ &   \multicolumn{4}{c |}{$0.5939 \le E $}\\
	\hline
	$K=7$ &  $0.5939 \le E  \le 0.6249 $ &   \multicolumn{3}{c |}{$1.744 \le E $}    \\%%%% Table body
	\hline 
	$K=8$ &  $0.6167 \le E  \le 0.6249 $ &   \multicolumn{3}{c |}{$1.744 \le E $}    \\%%%% Table body
	\hline 
	$K=9$ &  $0.6189 \le E  \le 0.6248 $ &   \multicolumn{3}{c |}{$1.983 \le E $}    \\%%%% Table body
	\hline 
	$K=10$ &  $0.6189 \le E  \le 0.6216 $ & $1.983 \le E  \le 2.034 $ &   \multicolumn{2}{c |}{$3.430 \le E $}    \\%%%% Table body
	\hline 
	$K=11$ &  $0.6207 \le E  \le 0.6216 $ & $2.021 \le E  \le 2.034 $ & $3.430 \le E  \le 4.109 $ &  $4.555 \le E $    \\%%%% Table body
	\hline 
	numerical & 0.6209 & 2.026 & 3.698 &  5.558   \\%%%% Table body
	\hline 
	\end{tabular}
	\caption{Energy spectra for the first four energy eigenstates ($n=1,2,3,4$) obtained by using the numerical bootstrap on $H=P^2/2+X^2/2+X^4/4$.
		We take $\{ X^n \}$, ($n=0,1,\cdots,K $) for constructing the bootstrap matrix in \eqref{ops-sample}.
	As $K$ increases, the constraint from the numerical bootstrap is getting stronger and the results become better.
	Besides, we obtain the energy eigenvalues from the lower eigenstates.
	For comparison, the results that are derived by solving the Schr$\ddot{\rm o}$dinger equation numerically are also shown at the last line in the table.
	}%%%Table caption goes here
	\label{Table-AHO}
	\end{table}%%%End of the table

	As $K$ increases, the results are getting better.
	For $K \le 6$, we obtain only the lower bound of energy, which is close to the ground state energy. 
	For $K \ge 7$, an isolated energy band that corresponds to the ground state appears.
	Similarly, at $K=10$, the band for the second energy eigenstate appears.
	These bands become narrower as $K$ increases and they tend to converge to the numerical results\footnote{In the numerical analysis, we numerically solve the Schr$\ddot{\rm o}$dinger equation with the Hamiltonian \eqref{H-AHO} by using Mathematica package NDEigensystem.} \footnote{
		The energy bands can be regarded as error bars of the bootstrap method.
		Note that one advantage of the bootstrap method is that these error bars  are exact \cite{Kazakov:2021lel}. Namely, the outside energy regions which are excluded by the bootstrap method are inconsistent with quantum mechanics, and they never survive.
		This is quite different from other numerical computations in which error bars are not exact.
		Thus, it would be valuable to investigate the numerical bootstrap method for large $K$ and compare the accuracy with other numerical method.
		}.

	Therefore, somehow the numerical bootstrap program works and we reproduce the correct energy eigen states. But one question is why it works. Another question is why we obtain the energy eigen states from the lower state. 
	In order to understand these questions, we study a harmonic oscillator in the next section.

\section{Bootstrap analysis in harmonic oscillator}
\label{sec-HO}

We study an one-dimensional harmonic oscillator,
\begin{align}
	H =& \frac{1}{2}\left( P^2+X^2 \right)   = a^\dagger a + \frac{1}{2},
	\label{HO}
\end{align}
where we have defined the ladder operators,
\begin{align}
	a= \frac{1}{\sqrt{2}}\left(
	X+iP
	\right), \quad
	a^\dagger = \frac{1}{\sqrt{2}}\left(
	X-iP
	\right).
	\label{ladder}
\end{align}
To employ the bootstrap method, we need to choose the set of operators \eqref{ops-sample}.
We find that the following operator is useful,
\begin{align}
	\tilde{O}_a:=  \sum_{n=0}^{K} c_{n} a^n
	=c_{0} +  c_{1} a  +  c_{2} a^2 + \cdots + c_{K} a^K .
	\label{ops-a}
\end{align}
Then, the bootstrap matrix \eqref{bootstrap} becomes,
\begin{align}
	{\mathcal M}_a:=
	\begin{pmatrix}
		1 & \left\langle a \right\rangle  & \left\langle a^2 \right\rangle   & \cdots & \left\langle a^K \right\rangle   \\
		\left\langle a^\dag	\right\rangle & \left\langle a^\dag a	\right\rangle & \left\langle a^\dag a^2 \right\rangle &  \cdots & \left\langle a^\dagger a^K \right\rangle   \\
		\left\langle (a^\dag)^2	\right\rangle  &  \left\langle (a^\dag)^2a \right\rangle & \left\langle (a^\dag)^2a^2 \right\rangle  & \cdots  &  \left\langle (a^\dag)^2a^K \right\rangle \\
		\vdots & \vdots  & \vdots  & \ddots & \vdots  \\
		\left\langle (a^\dag)^K  \right\rangle & \left\langle (a^\dag)^K a \right\rangle
		& \left\langle (a^\dag)^K a^2 \right\rangle & \cdots & \left\langle (a^\dag)^K a^K \right\rangle
	\end{pmatrix}.
	\label{bootstrap-A}
\end{align}

Now we consider the constraints \eqref{HO=0} and \eqref{HO=EO}.
By taking $O=(a^{\dagger})^m a^n  $  in these equations, we obtain
\begin{align}
	(m-n)  \langle (a^\dag)^m a^n \rangle=0, \qquad
	\langle (a^\dag)^{m+1} a^{n+1} \rangle = \left(E-m-\frac{1}{2}\right) \langle (a^\dag)^{m} a^{n} \rangle.
	\label{a-sol1}
\end{align}
Here, the first equation implies that $\langle (a^\dag)^m a^n \rangle=0$, if $m \neq n$.
Thus, the bootstrap matrix  ${\mathcal M}_a$ \eqref{bootstrap-A} becomes diagonal,
\begin{align}
	{\mathcal M}_a=
	\begin{pmatrix}
		1 & 0  & 0  & \cdots & 0   \\
		0 & \left\langle a^\dag a	\right\rangle & 0 &  \cdots & 0   \\
		0 & 0& \left\langle (a^\dag)^2a^2 \right\rangle  & \cdots  &  0 \\
		\vdots & \vdots  & \vdots  & \ddots & \vdots  \\
		0 & 0 & 0 & \cdots & \left\langle (a^\dag)^K a^K \right\rangle
	\end{pmatrix},
	\label{bootstrap2}
\end{align}
and the condition ${\mathcal M}_a \succeq 0$ reduces to 
\begin{align}
	\left\langle (a^\dag)^n a^n \right\rangle \ge 0, \qquad n=1,\cdots,K
	.
	\label{bootstrap-diagonal}
\end{align}
Besides, the second equation in \eqref{a-sol1} implies
\begin{align}
	\langle (a^\dag)^{n} a^{n} \rangle = 
	\prod_{k=0}^{n-1} \left(E-k-\frac{1}{2}\right) .
	\label{a-sol2}
\end{align}
Obviously, the conditions \eqref{bootstrap-diagonal} and \eqref{a-sol2} are equivalent to the Dirac's ladder operator problem and the solution for these conditions is given as
\begin{align}
	E=n+\frac{1}{2}, \quad \left( n=0,1,\cdots,K-2 \right), \quad \text{or} \quad E \ge K-\frac{1}{2}.
	\label{HO-solution}
\end{align}
Therefore, the bootstrap method reproduces the exact result when we take $K \to \infty$.

Our results may suggest the following answers for the two questions in the previous section.
\begin{itemize}
	\item The numerical bootstrap problem may be regarded as a numerical version of the Dirac's ladder operator approach. This may explain why the bootstrap method works in various systems.
	\item In \eqref{HO-solution}, for a given $K$, we obtain the eigenstates up to $(K-2)$-th level. This may explain why the bootstrap method derives the eigenstates from the lower states.
\end{itemize}

\subsection{Bootstrap analysis in Harmonic Oscillator via different bootstrap matrices}
\label{sec-numerics}

You may wonder that our derivation of the exact result is  a ``fine tuning" of the operators in the bootstrap matrix \eqref{bootstrap}, and, if we choose different operators from $a$ and $a^\dagger$, the result might change.
Indeed, the results depend on the choice of the operators.
Let us consider the following two sets of operators $\{ X^n \}$ and $\{  X^m P^n \}$.
In order to construct the corresponding two bootstrap matrices, we define
\begin{align}
	\tilde{O}_X :=  \sum_{n=0}^{K} c_{n} X^n, \qquad 
	\tilde{O}_{XP}:=  \sum_{m=0}^{K_X} \sum_{n=0}^{K_P} c_{mn} X^mP^n.
	\label{operators-XP}
\end{align}
Then, through the condition \eqref{positive-cond}, we obtain the bootstrap matrices
\begin{align}
	{\mathcal M}_X:=
	\begin{pmatrix}
		1 & \left\langle X \right\rangle  & \left\langle X^2 \right\rangle   & \cdots  \\
		\left\langle X	\right\rangle & \left\langle X^2	\right\rangle & \left\langle X^3 \right\rangle &  \cdots  \\
		\left\langle X^2	\right\rangle  &  \left\langle X^3 \right\rangle & \left\langle X^4 \right\rangle  & \cdots   \\
		\vdots & \vdots  & \vdots  & \ddots \\
	\end{pmatrix}, 
	\quad
	{\mathcal M}_{XP}:=
	\begin{pmatrix}
		1 & \left\langle X \right\rangle  & \left\langle P \right\rangle   & \cdots  \\
		\left\langle X	\right\rangle & \left\langle X^2	\right\rangle & \left\langle XP \right\rangle &  \cdots  \\
		\left\langle P	\right\rangle  &  \left\langle PX \right\rangle & \left\langle P^2 \right\rangle  & \cdots   \\
		\vdots & \vdots  & \vdots  & \ddots \\
	\end{pmatrix}.
	\label{bootstrap-XP}
\end{align}

Through the constraint equations \eqref{HO=0} and \eqref{HO=EO}, we obtain \eqref{recurrence-general} and \eqref{EO-XPn}.
These equations can be solved recursively and all the operator $\{  X^m P^n \}$ are described by energy $E$.
Then, the bootstrap matrices \eqref{bootstrap-XP} become,
\begin{align}
	{\mathcal M}_X:=
	\begin{pmatrix}
		1 & 0  &  E  & \cdots  \\
		0 &  E & 0 &  \cdots  \\
		E  &  0 & \frac{3}{2} E^2+ \frac{3}{8}   & \cdots   \\
		\vdots & \vdots  & \vdots  & \ddots \\
	\end{pmatrix}, 
	\quad
	{\mathcal M}_{XP}:=
	\begin{pmatrix}
		1 & 0  & 0   & \cdots  \\
		0 &  E & \frac{i}{2} &  \cdots  \\
		0  &  -\frac{i}{2} & E  & \cdots   \\
		\vdots & \vdots  & \vdots  & \ddots \\
	\end{pmatrix}.
	\label{bootstrap-XP2}
\end{align}

Now, we survey possible values of $E$ at which the bootstrap matrices become positive-semidefinite\footnote{
The bootstrap program in the harmonic oscillator is much easier than the anharmonic oscillator case.
On the borders of the parameter regions in which the bootstrap matrix is positive-semidefinite, some of the eigenvalues of the bootstrap matrix must become zero.
In the case of the harmonic oscillator, since the bootstrap matrices ${\mathcal M}_{X}$ and ${\mathcal M}_{XP}$ depend only on the single parameter $E$, we can easily find energies at which some eigenvalues become zero.
These energies are candidates for the upper or lower bounds of the energy bands, and we can easily test whether they are borders or not by evaluating the bootstrap matrices around these energies.
}. 
The results are summarized in Table \ref{Table-HO-X} and \ref{Table-HO-XP}.
We find that the exact energy eigenvalues are reproduced from ${\mathcal M}_{XP}$, while we cannot obtain the exact ones from ${\mathcal M}_{X}$.
In the case of ${\mathcal M}_{XP}$, since the ladder operators \eqref{ladder} can be constructed from the linear combinations of $X$ and $P$,
the bootstrap method may correctly capture the exact result discussed in the previous section.
Hence, we do not need a fine tuning.
Just considering both $P$ and $X$ in the bootstrap matrix would be enough to obtain the exact result.

\begin{table}
	\centering
	\begin{tabular}{|c||c|c|c|c|}%%%The number of columns has to be defined here
		\hline
		& $n= 1$   & $n= 2$  & $n= 3$  & $n= 4$ \\ %%%% Table body
		\hline
		\hline
		$K=5$ &   \multicolumn{4}{c |}{$0.4174 \le E $}\\
		\hline
		$K=6$ &  $0.4174 \le E  \le 0.5666 $ &   \multicolumn{3}{c |}{$1.131 \le E $}    \\%%%% Table body
		\hline 
		$K=9$ &  $0.4966 \le E  \le  0.5090 $ & $1.367 \le E  \le 1.779 $ &   \multicolumn{2}{c |}{$1.962 \le E $}    \\%%%% Table body
		\hline 
		$K=14$ &  $0.4999 \le E  \le  0.5001 $ & $1.497 \le E  \le 1.507 $ & $2.428 \le E  \le 2.566 $ &  $3.286 \le E $    \\%%%% Table body
		\hline 
		$K=15$ &  $0.4999 \le E  \le  0.5001 $ & $1.497 \le E  \le 1.502 $ & $2.474 \le E  \le 2.566 $ &  $3.286 \le E $    \\%%%% Table body
		\hline 
		exact & 0.5 & 1.5 & 2.5 &  3.5   \\%%%% Table body
		\hline 
	\end{tabular}
	\caption{Energy spectra for the first four eigenstates ($n=1,2,3,4$) obtained by using the numerical bootstrap on $H=P^2/2+X^2/2$.
		We take $\{ X^n \}$, ($n=0,1,\cdots,K $) in \eqref{operators-XP} for constructing the bootstrap matrix ${\mathcal M}_{X}$.
		At $K=6$, the first isolated energy band corresponding to the first energy eigenstate appears.
		At $K=9$ and $14$, the second and third energy bands appear, respectively.
		As $K$ increases, these bands are getting narrow, and tend to converge to the exact results.
		These properties are similar to the anharmonic oscillator case shown in Table \ref{Table-AHO}.
	}%%%Table caption goes here
	\label{Table-HO-X}
\end{table}%%%End of the table

\begin{table}
	\centering
	\begin{tabular}{|c||c|c|c|c|c|}%%%The number of columns has to be defined here
		\hline
		& $n= 1 $  & $n= 2 $ & $n= 3$  & $n= 4$ & $n= 5$ \\ %%%% Table body
		\hline
		\hline
		$K_X=1$, $K_P=1$ ($4 \times 4 $ matrix) &   \multicolumn{5}{c |}{$0.5 \le E $}\\
		\hline 
		$K_X=2$, $K_P=1$ ($6 \times 6 $ matrix) &  0.5 & \multicolumn{4}{c |}{$1.5 \le E $}\\
		\hline 
		$K_X=3$, $K_P=1$ ($8 \times 8 $ matrix) &  0.5 &  1.5 & \multicolumn{3}{c |}{$2.5 \le E $}\\
		\hline 
		$K_X=2$, $K_P=2$ ($9 \times 9 $ matrix) &  0.5 &  1.5 & \multicolumn{3}{c |}{$2.5 \le E $}\\
		\hline 
		$K_X=3$, $K_P=2$ ($12 \times 12 $ matrix) &  0.5 &  1.5 & 2.5 & \multicolumn{2}{c |}{$3.5 \le E $}\\
		\hline 
		$K_X=3$, $K_P=3$ ($16 \times 16 $ matrix) &  0.5 &  1.5 & 2.5 & 3.5  & $4.5 \le E $\\
		\hline 
		
		exact & 0.5 & 1.5 & 2.5 &  3.5 &  4.5   \\%%%% Table body
		\hline 
	\end{tabular}
	\caption{Energy spectra for the first five eigenstates ($n=1, \cdots , 5$) obtained by using the numerical bootstrap on $H=P^2/2+X^2/2$. We take $\{ X^m P^n \}$ in \eqref{operators-XP}, ($m=0,1,\cdots,K_X $ and $n=0,1,\cdots,K_P $) for constructing the bootstrap matrix ${\mathcal M}_{XP}$.
		Different from the ${\mathcal M}_{X}$ case shown in Table \ref{Table-HO-X}, the exact results are derived. 
	}%%%Table caption goes here
	\label{Table-HO-XP}
\end{table}%%%End of the table

On the other hand, in the case of ${\mathcal M}_{X}$, since we cannot construct the ladder operators only from $X$, the constraint ${\mathcal M}_{X} \succeq 0$ is not as strong as  ${\mathcal M}_{XP} \succeq 0$.
(If we take the size of the bootstrap matrix sufficiently large, the allowed region of $E$ are strongly constrained, and it becomes point-like, asymptotically \cite{Berenstein:2021dyf}.)

Note that we may claim that, if we compare two results obtained through two different bootstrap matrices whose sizes are the same, the method, which provides better results, is more efficient\footnote{This criterion for the efficiency of the bootstrap method is not so reasonable.
Rather, we should evaluate the efficiencies of different bootstrap methods by comparing their best results which are obtained at almost same numerical costs even with different sizes of  the bootstrap matrices.
However, the results would depend on numerical algorithms, and it is not easy to compare the efficiencies in this way. 
}. 
Let us compare the three $16\times 16$ bootstrap matrices: ${\mathcal M}_{X}$ at $K=15$, ${\mathcal M}_{a}$ at $K=15$ and ${\mathcal M}_{XP}$ at $K_X=K_P=3$.
There, the bootstrap method by using ${\mathcal M}_{a}$ is the best, since the 13 exact eigen states are derived in \eqref{HO-solution}, while the 4 exact states are derived from ${\mathcal M}_{XP}$ and the 3 states, which are not exact, are derived from ${\mathcal M}_{X}$ as shown in Table \ref{Table-HO-X} and \ref{Table-HO-XP}.

\subsection{Bootstraping the anharmonic oscillator via different bootstrap matrices}
\label{sec-AHO-2}

As we have seen in the previous subsection, the bootstrap matrix ${\mathcal M}_{a}$ \eqref{bootstrap-A} provides the best results in the harmonic oscillator.
Then, it is natural to ask whether ${\mathcal M}_{a}$ improves the numerical results in the anharmonic oscillator \eqref{H-AHO}.
In order to study this question, we rewrite the Hamiltonian \eqref{H-AHO} as
\begin{align}
	H= \omega  a^\dagger_\omega a_\omega +\frac{\omega}{2} + \frac{1}{2} \left(1 -\omega^2 \right) X^2 +\frac{1}{4}X^4, \qquad a_\omega := \sqrt{ \frac{\omega}{2} } X+i \frac{1}{\sqrt{2\omega}} P.
\end{align}
Here $\omega$ is a parameter for defining the ladder operator $a_\omega$, and the results would depend on $\omega$.
We simply take $\omega=1$, and numerically solve the bootstrap problem with the bootstrap matrix ${\mathcal M}_{a}$ \eqref{bootstrap-A}.
The results are shown in Table \ref{Table-AHO-a}.
We also perform the bootstrap analysis by using ${\mathcal M}_{XP}$ \eqref{bootstrap-XP} for comparison, and the result is summarized in Table \ref{Table-AHO-XP}.
Among the three $8 \times 8$ bootstrap matrices: ${\mathcal M}_{X}$ at  $K=7$, ${\mathcal M}_{a}$ at $K=7$ and ${\mathcal M}_{XP}$ at $K_X=3$, $K_P=1$, the results derived through ${\mathcal M}_{a}$ are the best, and ${\mathcal M}_{XP}$ is slightly better than ${\mathcal M}_{X}$.
Thus, the ladder operator might improve the bootstrap analysis generically.
% the bootstrap analysis in other systems also.

%Here, we have taken $K=8$ for  ${\mathcal M}_{X}$ and $K_x=K_p=2$ for ${\mathcal M}_{XP}$ in \eqref{operators-XP}. 
%Note that the sizes of the bootstrap matrix ${\mathcal M}_{X}$ and ${\mathcal M}_{XP}$ are both $9 \times 9$ in these cases.

\begin{table}
	\centering
	\begin{tabular}{|c||c|c|c|c|}%%%The number of columns has to be defined here
		\hline
		& $n=1$   & $n=2$  & $n=3$  & $n= 4$ \\ %%%% Table body
		\hline
		\hline
		$K=1$ &     \multicolumn{4}{c |}{NA}    \\%%%% Table body
		\hline 
		
		$K=2$ &  $0.4919\le E  \le 0.6767 $ &   \multicolumn{3}{c |}{$1.762 \le E $}    \\%%%% Table body
		\hline 
		$K=3$ &  $0.6041 \le E  \le 0.6314 $ &   \multicolumn{3}{c |}{$1.762\le E $}    \\%%%% Table body
		\hline 
		$K=4$ &  $0.6041 \le E  \le 0.6245 $ & $2.012 \le E  \le  2.454 $ &    \multicolumn{2}{c |}{$ 2.979 \le E $}    \\%%%% Table body
		\hline 
		$K=5$ &  $ 0.6194 \le E  \le 0.6245 $ & $2.012 \le E  \le  2.251 $ &    \multicolumn{2}{c |}{$  3.218 \le E $}    \\%%%% Table body
		\hline 
		$K=6$ &  $ 0.6194 \le E  \le 0.6212 $ & $2.020 \le E  \le  2.121 $ &    \multicolumn{2}{c |}{$  3.297 \le E $}    \\%%%% Table body
		\hline 
		$K=7$ &  $0.6207 \le E  \le 0.6212 $ & $2.020 \le E  \le 2.044 $ & $3.691 \le E  \le 4.099 $ &  $4.399 \le E $    \\%%%% Table body
		\hline 
%		$K=11$ &  $0.6209 \le E  \le 0.6210 $ & $2.025 \le E  \le 2.027 $ & $3.697 \le E  \le 3.702 $ &  $5.551 \le E $    \\%%%% Table body
%		\hline 
		numerical & 0.6209 & 2.026 & 3.698 &  5.558   \\%%%% Table body
		\hline 
	\end{tabular}
	\caption{Energy spectra for the first four eigenstates ($n=1, 2,3,4$) of the anharmonic oscillator $H=P^2/2+X^2/2+X^4/4$ via the bootstrap analysis with  the bootstrap matrix ${\mathcal M}_{a}$ constructed from the ladder operators $\{ a^m \}$.
	The results at $K=7$ is as good as $K=11$ in Table.\ref{Table-AHO}, which is obtained by using the operators $\{ X^m \}$.
	It might imply that the bootstrap method with the operators $\{ a^m \}$ is more efficient than the operators $\{ X^m \}$.
	}%%%Table caption goes here
	\label{Table-AHO-a}
\end{table}%%%End of the table

\begin{table}
	\centering
	\begin{tabular}{|c||c|c|c|c|}%%%The number of columns has to be defined here
		\hline
		& $n=1$   & $n=2$  & $n=3$  & $n= 4$ \\ %%%% Table body
		\hline
		\hline
		$K_X=1$, $K_P=1$ ($4 \times 4 $ matrix)  &      \multicolumn{4}{c |}{$0.4330 \le E$}    \\%%%% Table body
		\hline 
		$K_X=1$, $K_P=2$ ($6 \times 6 $ matrix)  &      \multicolumn{4}{c |}{$0.4916 \le E$}    \\%%%% Table body
\hline 
		$K_X=1$, $K_P=3$ ($8 \times 8 $ matrix)  &      \multicolumn{4}{c |}{$0.6188 \le E$}    \\%%%% Table body
\hline 		$K_X=2$, $K_P=1$ ($6 \times 6 $ matrix)  &      \multicolumn{4}{c |}{$0.6203 \le E$}    \\%%%% Table body
\hline 
		$K_X=3$, $K_P=1$ ($8 \times 8 $ matrix)  &   $0.6209 \le E  \le 0.6227$ &    \multicolumn{3}{c |}{$2.021\le E$}    \\%%%% Table body
\hline 
%$K_X=3$, $K_P=2$ ($12 \times 12 $ matrix)  &   $0.62092 \le E  \le 0.62094$ &
%$2.0259 \le E  \le 2.0261$ &  $3.698 \le E  \le 3.905$ &  $5.390 \le E$    \\%%%% Table body
%\hline 
%$K_X=3$, $K_P=2$  &   $0.6209 \le E  \le 0.6210$ &
%$2.025 \le E  \le 2.027$ &  $3.698 \le E  \le 3.905$ &  $5.390 \le E$    \\
%($12 \times 12 $ matrix)  &  & & &    \\%%%% Table body
%\hline 
%		$K_X=2$, $K_P=2$ ($9 \times 9 $ matrix)  &   $0.6209 \le E  \le 0.6212$ & $2.024 \le E  \le 2.620 $ &   \multicolumn{2}{c |}{$2.628\le E$}    \\%%%% Table body
%  \hline 
		numerical & 0.6209 & 2.026 & 3.698 &  5.558   \\%%%% Table body
		\hline 
	\end{tabular}
	\caption{
		Energy spectra of the anharmonic oscillator $H=P^2/2+X^2/2+X^4/4$ via the bootstrap analysis with the bootstrap matrix ${\mathcal M}_{XP}$ constructed from the operators $\{ X^mP^n \}$.	
%		The results are as good as the ones obtained from the bootstrap method of $\{ X^m \}$ shown in Table \ref{Table-AHO}, and the results through  $\{ a^m \}$ shown in Table \ref{Table-AHO-a} is better.
	}%%%Table caption goes here
	\label{Table-AHO-XP}
\end{table}%%%End of the table

\section{Discussion}
\label{sec-discussion}

In this article, we have shown that the bootstrap problem in the harmonic oscillator reduces to the Dirac's ladder operator problem and exactly solvable. 
We have also demonstrated that the ladder operator improves the bootstrap analysis in the anharmonic oscillator \eqref{H-AHO}.
These results suggest the connection between the bootstrap problem and the Dirac's ladder operator problem.
Thus, the bootstrap method may be regarded as a numerical version of the Dirac's ladder operator problem, and it may explain why this method works in various systems.

Actually, we can derive the ground state energy $E=N^2/2$ in the $U(N)$ matrix quantum mechanics with a harmonic potential,
\begin{align}
	H=N \Tr\left( \frac{1}{2}P^2+\frac{1}{2}X^2 \right),
	\label{MQM}
\end{align}
through the bootstrap method similar to our ladder operator analysis discussed in Sec.~\ref{sec-HO}.
Here, $X$ is an $U(N)$ hermitian matrix and $P$ is its conjugate momentum that satisfies the commutator relation $[P_{ij},X_{kl}]=-i \delta_{il} \delta_{jk}/N$, $(i,j,k,l=1,\cdots,N)$.
Thus, the bootstrap problem in many body systems may also be explained in terms of the  ladder operators.

\section*{Acknowledgment}

The authors would like to thank Takehiro Azuma for valuable discussions and comments.
A part of numerical computation in this work was carried out at the Yukawa Institute Computer Facility.
The work of T.~M. is supported in part by Grant-in-Aid for Scientific Research C (No. 20K03946) from JSPS.

\appendix

\section{Solving the constraints \eqref{HO=0} and \eqref{HO=EO}}
\label{app-recurrence}

In this appendix, we show how to solve the constraint equations \eqref{HO=0} and \eqref{HO=EO} when the Hamiltonian is given by
\begin{align}
	H=\frac{1}{2}P^2+V(X),
\end{align}
where $V(X)$ is a polynomial of $X$.
First we take $O=X^n$ in \eqref{HO=0} and \eqref{HO=EO}, and obtain
\begin{align}
	\langle [H,X^n] \rangle =0 & \Rightarrow  \langle X^{n-1} P \rangle =  \frac{i}{2}(n-1) 	\langle X^{n-2} \rangle  ,
	\label{HO-X}
	\\
		\langle X^n H \rangle =E\langle X^n  \rangle  & \Rightarrow 
 \langle X^{n} P^2 \rangle = -2 \langle X^n  V(X) \rangle +2 E \langle X^n  \rangle .
 \label{EO-X}
\end{align}
Similarly, by taking $O=X^nP$, we obtain
\begin{align}
	\langle [H,X^n P] \rangle =0 & \Rightarrow 
	  -n(n-1) \langle X^{n-2} P \rangle -2in \langle X^{n-1} P^2 \rangle +2i \langle X^{n} V'(X) \rangle =0.
	  \label{HO-XP} \\
	\langle X^n P H \rangle =E 	\langle X^n P  \rangle & \Rightarrow 
	 \langle X^{n} P^3 \rangle  = 2E \langle X^{n} P \rangle +2i \langle X^{n} V'(X) \rangle -2\langle X^{n} V(X)P \rangle .
	\label{EO-XP} 
\end{align}
By substituting $ \langle X^{n-2} P \rangle$ in \eqref{HO-X} and $\langle X^{n-1} P^2 \rangle$ in \eqref{EO-X} into \eqref{HO-XP}, we obtain a recurrence relation \cite{Berenstein:2021dyf, Bhattacharya:2021btd}
\begin{align}
	n(n-1)(n-2) \langle X^{n-3} \rangle -8n \langle X^{n-1} V(X)\rangle +8 n E \langle X^{n-1}  \rangle -4 \langle X^{n} V'(X)\rangle=0.
	\label{recurrence-general}
\end{align}
Since $V(X)$ is polynomial, we can solve this relation and $ \langle X^{n} \rangle $ would be described by a set of the expectation values $\{ \langle X^{l} \rangle \} $ and $E$.
In Sec.\ref{sec-review}, we use this recurrence relation to obtain \eqref{AHO-recurrence}.

We can show that the operators involving momentum $\{ \langle X^{m} P^n \rangle \} $ are also described by  $\{ \langle X^{l} \rangle \} $ and $E$.
Here we consider only the operators $ \langle X^{m} P^n \rangle  $ because the operators  $ P^n X^{m}  $  
can be written as
\begin{align}
	P^n X^{m} =
	\sum_{k=0}^{\min (m,n)}  (-i)^{k} \frac{n!m!}{k!(n-k)!(m-k)!}  X^{m-k} P^{n-k}
	\label{ordering}
\end{align}
through the commutation relations, and they are described by the ordered operators $ \{ X^{a}  P^b \} $, ($0 \le a\le m$, $0 \le  b \le n$).

Indeed, from \eqref{HO-X}, \eqref{EO-X} and \eqref{HO-XP},
 the operators $\langle X^{m} P \rangle$, $\langle X^{m} P^2 \rangle$ and $\langle X^{m} P^3 \rangle$ are described by $\{ \langle X^{l} \rangle \} $ and $E$.
Similarly, we obtain the relations for $\langle X^{m} P^n \rangle$ by taking $O=X^mP^{n-2}$ in \eqref{HO=EO}, 
\begin{align}
	\langle X^m P^{n-2} H \rangle =E 	\langle X^m P^{n-2}  \rangle & \Rightarrow 
	\langle X^{m} P^n\rangle  = 2E \langle X^{m
	} P^{n-2} \rangle -2 \langle X^{m} P^{n-2} V(X) \rangle .
	\label{EO-XPn} 
\end{align}
Since $\langle X^{m} P^{n-2} V(X) \rangle $ can be described by operators $\langle X^a P^b \rangle $, ($b \le n-2$) through \eqref{ordering}, we can solve $\langle X^{m} P^n \rangle$ from lower $n$ recurrently.
In Sec.~\ref{sec-numerics}, we use this relation in \eqref{bootstrap-XP2}.

Note that we have not used the constraint \eqref{HO=0} with $O=X^mP^n$, ($n \ge 2 $), and you may wonder if they provide additional constraints.
Actually, the answer is no.
This can be shown by induction.
Assuming that \eqref{HO=0} and \eqref{HO=EO} with respect to $O=X^mP^k$, $(0\le k \le n-1)$ are solved, and $\langle X^mP^k \rangle$ for all $m$ and $k$ up to $n+1$ are described by $\{ \langle X^{l} \rangle \} $ and $E$. 
Then, from \eqref{HO=0} with $O=X^mP^n$, we obtain
\begin{align}
	\langle [H,X^m P^n] \rangle & =
	\frac{1}{2} \langle [P^2,X^m ]P^n \rangle  + \langle X^m[V(X),P^n] \rangle  \nonumber \\
&	=
	\frac{1}{2} \langle [P^2,X^m ]P^n \rangle + \langle X^m[V(X),P^{n-2}]P^2 \rangle  + \langle X^mP^{n-2}[V(X),P^2] \rangle  
	\label{HO-XPn} 
\end{align}
Here, from \eqref{HO=0} with $O=X^m  P^{n-2} V(X) $, we obtain\footnote{By using \eqref{ordering}, $\langle [H,X^m  P^{n-2} V(X) ] \rangle$ can be written as a sum of $\langle [H,X^a  P^{b}  ] \rangle$ with $b \le n-2$.
Thus this equation is zero from the assumption of the induction.}
\begin{align}
	0&=\langle [H,X^m  P^{n-2} V(X) ] \rangle \nonumber \\
	& =
	\frac{1}{2} \langle  [P^2, X^m  ]  P^{n-2} V(X)  \rangle  + 
	\frac{1}{2} \langle  X^m  P^{n-2} [P^2, V(X) ]    \rangle +
	\langle  X^m     [V(X),  P^{n-2}] V(X)  \rangle .
	\label{HO-XVP}
\end{align}
By using this relation, we eliminate $ \langle X^mP^{n-2}[V(X),P^2] \rangle $ in \eqref{HO-XPn} and obtain
\begin{align}
&	\langle [P^2,X^m ]P^{n-2} \left( \frac{1}{2} P^2+V(X) \right) \rangle + \langle 2X^m[V(X),P^{n-2}] \left( \frac{1}{2} P^2 +V(x)  \right)   \rangle    \nonumber \\
=& \langle [P^2,X^m ]P^{n-2} H \rangle + \langle 2X^m[V(X),P^{n-2}] H  \rangle  
=2E \langle \left[\frac{1}{2} P^2+V(X), X^m P^{n-2} \right]  \rangle =0
\end{align}
Here we have used $\langle X^a  P^b H \rangle = E \langle X^a  P^b \rangle $ and $\langle [H, X^a  P^b ] \rangle =0$ with $b \le n-1$. 
Thus, the constraint \eqref{HO=0} with  $O=X^mP^n$, ($n \ge 2 $) is automatically satisfied, and no additional constraint appears.

{\normalsize 
\bibliographystyle{unsrt}
 \bibliography{bBFSS} }

\end{document}